# Tunable terahertz plasmons in graphite thin films

Qiaoxia Xing[1,2], Chaoyu Song[1,2], Chong Wang[1,2], Yuangang Xie[1,2], Shenyang Huang[1,2], Fanjie Wang[1,2], Yuchen Lei[1,2], Xiang Yuan[1,6], Cheng Zhang[1,3], Lei Mu[1,2], Yuan Huang[5], Faxian Xiu[1,3,4] and Hugen Yan[1,2]*

[1]State Key Laboratory of Surface Physics, Department of Physics, Fudan University, Shanghai 200433, China.

[2]Key Laboratory of Micro and Nano-Photonic Structures (Ministry of Education), Fudan University, Shanghai 200433, China.

[3]Institute for Nanoelectronic Devices and Quantum Computing, Fudan University, Shanghai 200433, China.

[4]Shanghai Research Center for Quantum Sciences, Shanghai 201315, China

[5]Institute of Physics, Chinese Academy of Sciences, 100190 Beijing, China.

[6]State Key Laboratory of Precision Spectroscopy, East China Normal University, Shanghai 200062, China.

*Corresponding author: hgyan@fudan.edu.cn (H. Y.)

**Tunable terahertz plasmons are essential for reconfigurable photonics, which have been demonstrated in graphene through gating, though with relatively weak responses. Here, we demonstrate strong terahertz plasmons in graphite thin films via infrared spectroscopy, with dramatic tunability by even a moderate temperature change or an in-situ bias voltage. Meanwhile, through magneto-plasmon studies, we reveal that massive electrons and massless Dirac holes make comparable contributions to the plasmon response. Our study not only sets up a platform for further exploration of two-component plasmas, but also opens an avenue for terahertz modulation through electrical bias or all-optical means.**

Tunable terahertz photonic devices are indispensable in terahertz applications in sensing, imaging, waveguiding etc. [1-5]. Various tunable plasmonic materials operating in the terahertz range have been explored, including graphene [6, 7], phase-changing compounds [8-11] and carbon nanotubes [12-14]. In particular, with good gate tunability, graphene is very promising for terahertz plasmonic applications [15, 16]. However, due to relatively weak response to light [17], graphene has to be combined with metallic structures to realized feasible modulation [18-24]. An extension to graphite thin films beyond monolayer is a natural avenue to achieve stronger and more intrinsic plasmonic response. Though the gate electrical field is screened in graphite thin films, the thermal carrier density depends strongly on temperature [25, 26], which promises sensitive tuning of plasmons by temperature.

Compared to phase-changing compounds, which typically exhibit switchable plasmons around the critical temperature [9], the plasmons in graphite are expected to be tuned continuously in a broad temperature range. Thermo-tuning of plasmons, in particular with ultrafast lasers to excite carriers [27-29], promises applications in all-optical modulation and tunable plasmonic metamaterials.

Moreover, graphite is a semimetal where massive electrons and massless Dirac holes coexist, residing around K-point and H-point of the Brillouin zone, respectively [30-32], and forming a two-component plasma [33-35]. Depending on their relative oscillation phase, the collective oscillations of electrons and holes can be categorized into optical and acoustic modes [34, 36]. The interrogation of plasmons in graphite thin films can possibly gain insight into the many-body interaction in the two-component plasma, which involves both regular and massless Dirac fermions.

In our work, we perform a systematic study of the graphite plasmon in the terahertz regime using far-field infrared spectroscopy. Strong temperature and bias voltage dependence of the plasmon has been revealed. The magnetic field effect on the plasmon differentiates the contributions of massive and massless fermions to the collective oscillation and comparable Drude weights from both components are inferred.

A schematic of the far-infrared transmission measurement is shown in Fig. 1(a). Extinction spectra $1-T/T_0$ characterizes the electromagnetic responses of graphite thin films or microstructure arrays. A typical exfoliated graphite thin film on Si substrate is displayed in Fig. 1(b). Details of the fabrication and measurement procedures are given in Supplemental Material [37]. The extinction spectra of a typical unpatterned graphite film with the thickness of ~20 nm are shown in Fig. 1(c). At liquid helium cryogenic temperatures, the free carrier Drude response is still quite pronounced even without thermal excitation, and a step-like feature, indicating the onset of interband transitions, is observed around 400 cm$^{-1}$. These behaviors are consistent with the band structure of graphite [31], in which the hole pocket with a linear dispersion near the H point coexists with an electron pocket with a parabolic dispersion near K point [30], and the Fermi level determines the onset of interband transitions, as illustrated in the inset of Fig. 1(c).

Due to the semimetal nature and the Fermi energy in a close proximity of the band touching in graphite, the Drude response and interband excitations are sensitive to temperature. Thermal excitation can efficiently increase the free carrier density, resulting in an enhancement of the Drude response and suppression of interband transitions due to Pauli blocking. Such blocking prohibits the optical transitions from empty states in the valence band or into filling states in the conduction band, which have been created by thermal excitation. We plot the extracted Drude weight (see Supplementary Material [37] for details) as a function of temperature in Fig. 1(d) which is consistent with the prediction for 2D electron gas with touched valence and conduction bands:

$$D = Ak_B T \ln\left(2\cosh\left(\frac{\mu}{2k_B T}\right)\right) \quad (1)$$

where $D$ is the Drude weight, $A$ is a fitting coefficient, $T$ is the temperature, $k_B$ is Boltzmann constant, and $\mu$ is the chemical potential. For simplicity, we treat $\mu = E_F(T = 0 \text{ K})$ as a temperature-independent parameter, which is an accurate exercise for parabolic bands (see Supplementary Material [37] for details) [40, 41]. The fitting value of the chemical potential is 21 meV, in good agreement with the literature [31].

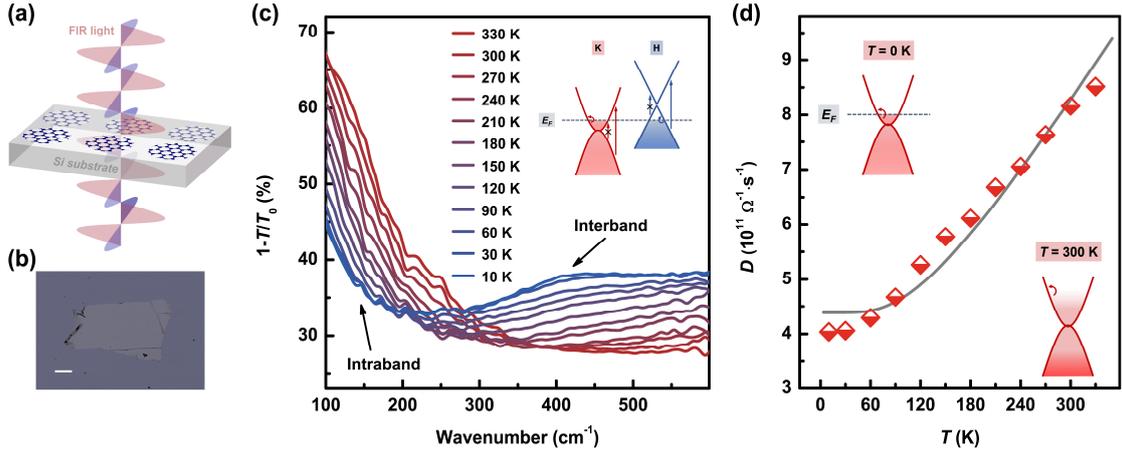

FIG. 1. Characterization of the exfoliated graphite thin film. (a) An illustration of the far-infrared spectroscopy scheme. (b) A typical optical image of the exfoliated graphite thin film on Si substrate, scale bar 100 μm. (c) Temperature-dependent far-infrared spectra of a graphite film with thickness of ~20 nm. The inset shows the electron (K point) and hole (H point) pockets in the Brillouin zone at zero temperature. The arrows (with a cross) indicate intraband and (Pauli-blocked) interband transitions, and the gray dashed line represents the Fermi energy. (d) Drude weight in (c) as a function of temperature. The curve is the fitting based on Eq. (1). The inserted sketches illustrate the carrier distributions at zero and room temperature, and the arrows represent the intraband transitions.

The temperature dependence of the Drude weight for graphite films is inherited by plasmons. We patterned a continuous thin film with thickness of 15 nm into an array of disks with diameter of 2.3 μm, as shown by the scanning electron microscope (SEM) image in the inset of Fig. 2(a). With the temperature increasing from 5 to 330 K, the plasmon frequency increases 60%, and the intensity increases to 2 times of the original accordingly, as demonstrated in Fig. 2(a). The extinction of the plasmon in graphite films is generally much larger than that in graphene, and the linewidth is comparable (see Supplementary Material [37]), which are desirable for real applications. In addition, the field confinement factor is ~15 at 5 K, much better than that even for the thinnest noble metal structures in IR frequencies [42]. Figure 2(b) shows the plasmon frequency and spectrum weight (procedures to extract the weight are in Supplementary Material [37]) as functions of temperature. In the low frequency (long wavelength) regime, the two-dimensional (2D) plasmon frequency scales as $\sqrt{D}$, where $D$ is the Drude weight [6, 15, 43]. Such a scaling works very well for the plasmon in the graphite thin film, as shown in Fig. 2(c), where we plot the plasmon frequency in Fig. 2(b) as a function of Drude weight $D$ at each temperature obtained from a graphite film with a similar thickness.

The temperature dependence is so sensitive that even the Joule heating of an electrical device based on the graphite thin film can largely in-situ modulate the plasmonic response. As demonstrated in Fig. 2(d), a current passing through a graphite ribbon array induces a pronounced shift of the plasmon frequency and an enhancement of the intensity, with the device both at low and room temperature environments. This promises reconfigurable metasurfaces through in-situ bias-tuning.

In addition to the graphite thin film, for comparison, we also performed a temperature dependent measurement of a 4-layer graphene disk array on $SiO_2$/Si substrate (mid-IR spectrum of the sample is in Supplementary Material [37]). As shown in Fig. 2(e), the plasmon frequency and intensity are almost independent of the temperature, since few-layer graphene tends to be doped to a high Fermi level (equivalent to thousands of Kelvin) by the surrounding media, and according to Eq. (1), the Drude weight is almost a constant within our temperature range. The temperature insensitivity of highly doped graphene is in marked contrast to that of graphite thin films with a much lower Fermi level.

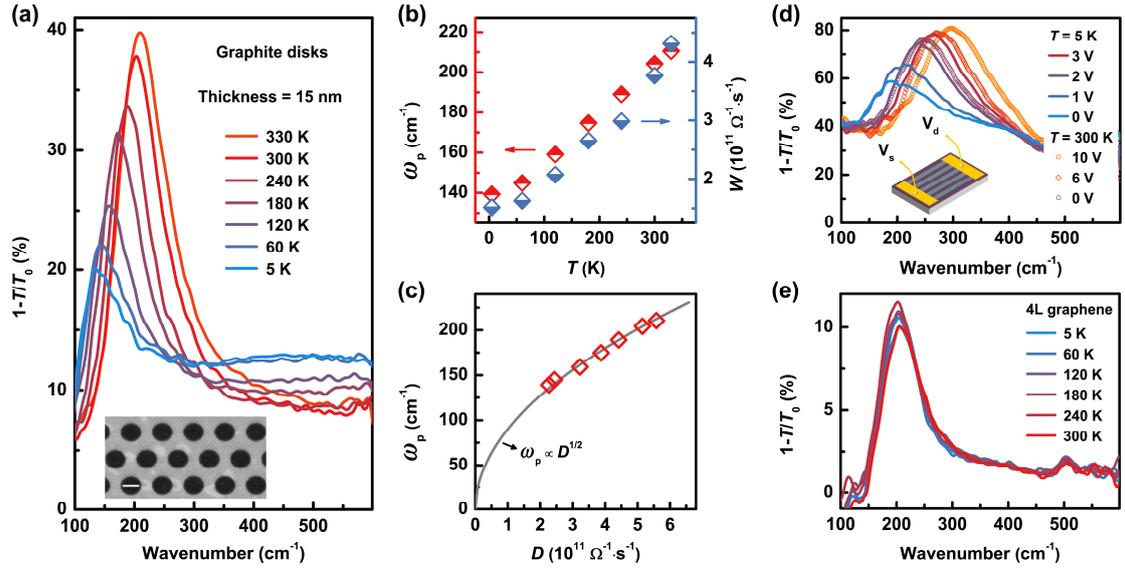

FIG. 2. The temperature-dependent plasmon. (a) Temperature-dependent far-infrared spectra of a graphite disk array with the diameter of 2.3 μm and the period of 3.75 μm. The inset shows the SEM image, scale bar 2 μm. The graphite film thickness is ~15 nm. (b) Plasmon frequency and weight $W$ extracted from (a) as functions of temperature $T$. (c) Plasmon frequency as a function of the Drude weight $D$ extracted from a graphite film with a similar thickness as that of the disk array. (d) Far-infrared

spectra of a biased graphite ribbon array with the ribbon width of 4 μm and thickness of ~30 nm at 5 K and room temperature. The inset is an illustration of a typical device. (e) Temperature-dependent far-infrared spectra of a 4-layer graphene disk array with disk diameter of 1 μm.

By shrinking the structure size, the plasmon can be pushed to higher frequencies. A series of graphite micro-rectangle arrays from a large graphite film with thickness of 15 nm was fabricated. The inset of Fig. 3(a) shows the typical morphology of the graphite rectangle array. As shown in Fig. 3(a), with the decreasing of the side length from 3 to 0.5 μm, the plasmon blueshifts, and its intensity decreases. At the same time, the fitted linewidth of the plasmon gradually increases from 90 cm$^{-1}$ to 203 cm$^{-1}$, as displayed in Fig. 3(b). The plasmon broadening and intensity reduction are compelling evidences of the increasing Landau damping, since the interband transition channels start to dominate beyond ~400 cm$^{-1}$, making the annihilation of higher energy plasmons into electron-hole pairs efficient. Nevertheless, as a two dimensional film, the plasmon dispersion still follows the standard $\omega_p \propto \sqrt{q}$ scaling law [43], as verified in Fig. 3(c), where $q = \pi/L$ ($L$ is the rectangle side length) is the plasmon wave vector, without considering the plasmon phase shift at boundaries for simplicity [45]. The plasmon dispersion, broadening and intensity can also be confirmed by the loss function [46, 47] $-\text{Im}(1/\varepsilon)$, as displayed as a pseudocolor map in Fig. 3(c). The calculation details are presented in the Supplementary Material [37]. As we can see from the map, the plasmon broadens with increasing frequency and beyond certain frequency (wave vector), the plasmon peak is almost completely smeared out. Such cut-off frequency depends on the temperature and higher temperature can sustain higher cut-off frequency due to the reduced interband Landau damping originated from the band-filling effect [26, 50], as exemplified by the simulation of a graphite rectangle array and the response of graphite split rings in Supplementary Material [37].

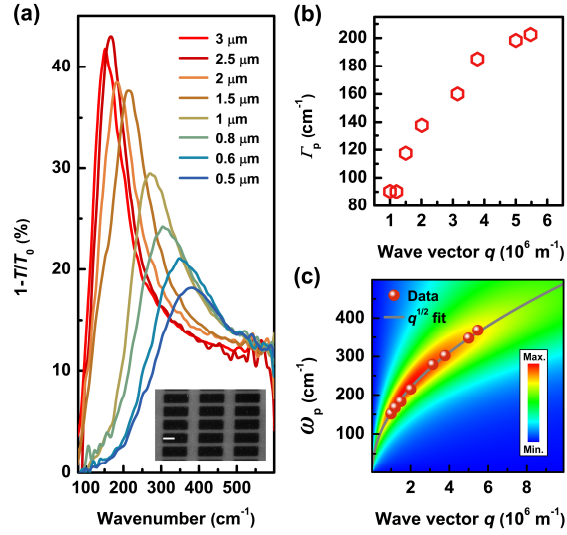

FIG. 3. Plasmon dispersion. (a) Extinction spectra of rectangle arrays with decreasing side length. The thickness of the graphite film is about 15 nm. The inset shows an SEM image of a typical rectangle array with a scale bar of 1 μm. (b) Plasmon linewidth as a function of the wave vector $q$. (c) Plasmon frequency in (a) as a function of the wave vector. The grey curve is a $q^{1/2}$ scaling. The loss function is also shown as a background.

The response of graphite plasmon to an external magnetic field is of potential interest due to the small cyclotron mass [54]. Moreover, owing to very different nature of carriers in the H- and K-pockets, the free carriers in graphite can be treated as a two-component plasma and a magnetic field can quantify their contributions to the collective excitation. In a disk with a single component carrier, the magneto-plasmon resonances have two modes, the bulk mode $\omega^+$ and the edge mode $\omega^-$ [55]:

$$\omega^\pm = \sqrt{\omega_0^2 + \left(\frac{\omega_c}{2}\right)^2} \pm \frac{\omega_c}{2} \qquad (2)$$

where $\omega_0$ is the plasmon frequency at zero magnetic field, $\omega_c$ is the cyclotron frequency. For massive carriers, $\omega_c = eB/m_c$, with $m_c$ as the cyclotron mass. In graphite disks, since Dirac holes at H-point exhibit much larger cyclotron frequency than massive electrons at K-point, based on Eq. (2), the originally coupled electron-hole plasma will be decoupled even at a moderate magnetic field. Therefore, we expect to see magneto-plasmon modes contributed solely by one of the carrier components.

We performed magneto-transmission measurements with the magnetic field perpendicular to the graphite disk array (Faraday configuration). The extinction spectrum without an external magnetic field is shown in Fig. 4(a), with a peak frequency of ~135 cm$^{-1}$ at 10 K. The measured relative transmission spectra with the magnetic field ranging from 0 T to 17.5 T are presented in Fig. 4(b). The sample was always at the liquid-helium temperature, so the thermal excitation is minimal. The possible excitonic effect can be ignored due to the only moderate magnetic field in our study [56]. With the increase of the field, the frequency and magnitude of the dip indicated by the orange arrow dramatically increase. Other weaker dips are higher energy Landau level transitions of electrons, as detailed in Supplementary Material [37]. The features of massless Dirac holes at H-point [57, 58], which typically have much higher frequencies and scale as $\sqrt{B}$, are not clearly observed. We plot the extracted frequency of the major dip as a function of the magnetic field ($B$ not less than 4T) in Fig. 4(c) and fit it with Eq. (2). The fitted cyclotron mass $m_c \approx 0.039 m_0$ ($m_0$ is the free electron mass) is consistent with that in the literature for K-pocket electrons [58]. As expected, the fitted plasmon frequency $\omega_0 \approx 93 \text{ cm}^{-1}$ is smaller than the measured total frequency (135 cm$^{-1}$) of the two-component carriers at zero field. These behaviors clearly suggest that the electrons and holes are decoupled under $B$-field due to the unparalleled cyclotron frequency (See Fig. 4(c) for the cyclotron frequencies of holes and electrons, and the magneto-transmission spectra of an unpatterned graphite thin film shown in Supplementary Material [37].), and the observed magneto-plasmon mode is from a single component, i.e., electrons in the K-pocket. More quantitatively, the total Drude weight $D_{total} = D_e + D_h$, with $D_e$ and $D_h$ as Drude weights from electrons in K-pocket and holes in H-pocket at zero field, respectively. Because the plasmon frequency $\omega \propto \sqrt{D}$, we have $\omega_{total}^2 = \omega_e^2 + \omega_h^2$. With the aforementioned $\omega_{total}$ and $\omega_e = \omega_0$, we get $\omega_h \approx 97 \text{ cm}^{-1}$. Therefore, the Drude weights of electrons and holes in graphite are comparable, which is consistent with previous results [59, 60]. The lower branch $\omega^-$ of the split plasmon (edge mode [55]) in Eq. (2) is not observed in our experiment, presumably due to its resonance frequency beyond our lower measurement limit (about 80 cm$^{-1}$) after both branches are well-split. Here we plot it based on Eq. (2) with the obtained parameters, as shown by the orange dashed line in Fig. 4(c).

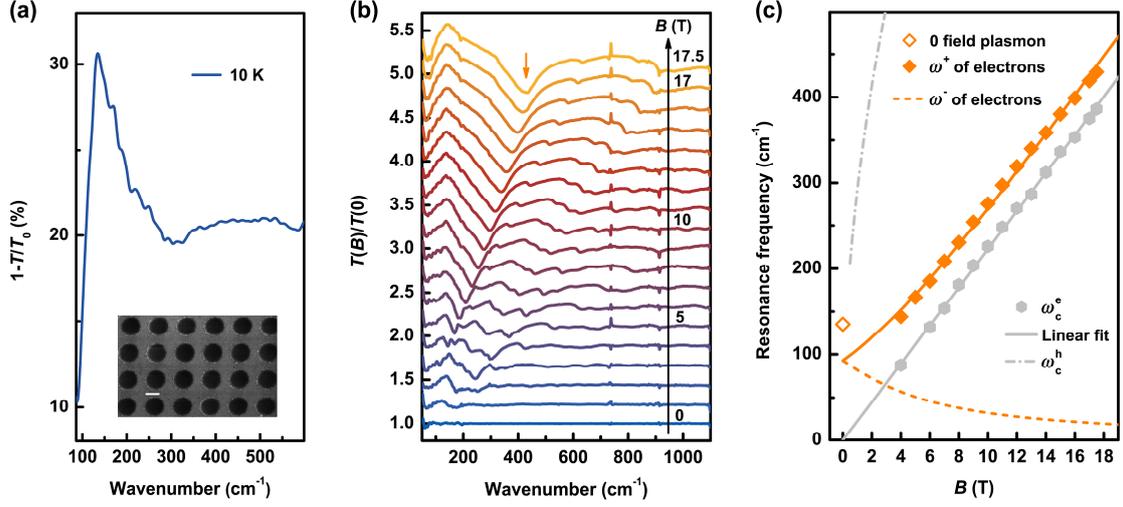

FIG. 4. Magneto-plasmon. (a) Extinction spectra of a disk array with the diameter of 3 μm at 10 K without a magnetic field. The inset shows the SEM image, scale bar 2 μm. The graphite film thickness is ~20 nm. (b) Relative transmission spectra $T(B)/T(0)$ in a Faraday configuration for the sample in (a). For clarity, spectra are shifted vertically. (c) Extracted frequency (orange data points) of the dip indicated by the orange arrow in (b) as a function of the magnetic field. The orange solid line is the fitting curve for the bulk mode based on Eq. (2), and the edge mode is also plotted as an orange dashed line. Zero field plasmon frequency determined in (a) is also shown for comparison. The light gray dashed line is the cyclotron frequency of holes in H-pocket, which is plotted based on the documented Fermi velocity of the Dirac cone [58]. The light gray data points are the cyclotron frequencies of electrons in K-pocket, extracted from the spectra in Supplementary Material [37]. The light gray solid line is the linear fitting.

In conclusion, we have systemically studied the tunable plasmon in graphite thin films through infrared spectroscopy. In addition to strong optical response, graphite plasmon also manifests pronounced temperature and bias voltage dependence, which is in sharp contrast to the plasmon in graphene. The response of the plasmon mode in a magnetic field suggests that electrons and holes contribute similar Drude weights. In addition to the temperature and a magnetic field, plasmons in graphite thin films can be potentially tuned by intercalation as well [61], which may uncover rich physics and unleash huge potential for applications.

## Acknowledgments

H. Y. is grateful to the financial support from National Natural Science Foundation of China (Grants No. 12074085, No. 11734007), the National Key Research and Development Program of China (Grants No. 2016YFA0203900 and No. 2017YFA0303504), the Strategic Priority Research Program of Chinese Academy of Sciences (XDB30000000), and the Natural Science Foundation of Shanghai (Grant No. 20JC1414601). C. W. is grateful to the financial support from the National Natural Science Foundation of China (Grant No. 11704075) and China Postdoctoral Science Foundation. F. X. is grateful to the financial support from the National Key Research and Development Program of China (Grants No. 2017YFA0303302 and No. 2018YFA0305601), the National Natural Science Foundation of China (Grants No. 11934005, No. 61322407, No. 11874116, No. 61674040), the Science and Technology Commission of Shanghai (Grant No. 19511120500), the Program of Shanghai

Academic/Technology Research Leader (Grant No. 20XD1400200), and the Shanghai Municipal Science and Technology Major Project (Grant No. 2019SHZDZX01). Y. H. is grateful to the financial support from National Natural Science Foundation of China (Grant No. 11874405), the National Key Research and Development Program of China (Grant No. 2019YFA0308000) and Strategic Priority Research Program of Chinese Academy of Sciences (XDB33000000). Part of the experimental work was carried out in the Fudan Nanofabrication Lab. Magneto-optical measurements were performed at the National High Magnetic Field Laboratory, which is supported by the National Science Foundation through NSF/DMR-1157490 and DMR-1644779 and the State of Florida. The authors thank Z. Z. Sun and W. Li for the help in measuring the thickness of the samples.